\documentclass[10pt,a4paper]{article}
\usepackage[T1]{fontenc}
\usepackage[left=36.85bp,right=36.85bp,top=34bp,bottom=63bp]{geometry}
\usepackage{amsmath,txfonts,graphicx,multicol}
\usepackage[scaled=0.94]{helvet}
\usepackage{microtype}
\usepackage[authoryear,round]{natbib}
\setcitestyle{aysep={}}
\usepackage{caption}
\usepackage{xcolor}
\usepackage[colorlinks=true,allcolors=blue,breaklinks=true]{hyperref}
\usepackage{orcidlink}
\DeclareMathSizes{10}{10}{5.8}{4}
\DeclareMathSizes{9}{9}{5.3}{4}
\newcommand{\papertitle}{Detection of GRB 261010A with GRID-11B}
\title{\papertitle}
\author{Chenyu Wang, Longhao Li, Xiaofan Pan, Hua Feng, Ming Zeng, and Zihan Yu}
\date{16 November 2026}
\hypersetup{pdftitle={\papertitle},pdfauthor={Chenyu Wang, Longhao Li, Xiaofan Pan, Hua Feng, Ming Zeng, Zihan Yu},pdfkeywords={Gamma-ray bursts; GRB 261010A; Prompt emission}}
\newcommand{\journal}{Progress in High Energy Astrophysics}
\newcommand{\pageid}{page~\thepage} 
\newcommand{\sizebp}[2]{\fontsize{#1bp}{#2bp}\selectfont}
\newcommand{\abstracttext}{We report GRID-11B observations of the main emission episode of GRB~261010A. We measure $T_{90}=5.68^{+1.04}_{-0.32}$~s and a minimum variability timescale of $51.6^{+10.5}_{-9.8}$~ms. Its spectrum is described by a cutoff power law with $E_{\rm p}=1089^{+171}_{-80}$~keV. At both candidate host redshifts ($z=0.070$ and $0.785$), the low $\varepsilon$ values and their positions relative to the Amati relation suggest Type-I-like properties. These properties, despite the long duration, resemble those of GRB~211211A and motivate investigation of a possible compact-object merger origin.}
\newcommand{\keywords}{Gamma-ray bursts; GRB~261010A; Prompt emission}

\DeclareCaptionFont{published}{\fontsize{10}{10.1bp}\selectfont}
\begin{document}
{\sizebp{10}{11}\noindent\textit{\journal}\enspace|\enspace\textsf{\textbf{PROMPT DISCOVERY}}\hfill\textcopyright~The Author(s) 2026\par
\noindent November 2026, 1(1)\par
\noindent\href{https://doi.org/10.3724/phea-20260001}{https://doi.org/10.3724/phea-20260001}\par
\noindent\href{https://cstr.cn/32526.14.phea.20260001}{https://cstr.cn/32526.14.phea.20260001}\par}
\vspace{8bp}
{\centering\sffamily\bfseries\sizebp{17.5}{19}\papertitle\par}
\vspace{24bp}
\noindent\hfill\begin{minipage}{\dimexpr\textwidth-46bp\relax}
\sizebp{9}{10.3}\textsf{\textbf{ABSTRACT:}} \abstracttext\ \textsf{\textbf{Key words:}} \keywords
\end{minipage}\hfill\null\par
\vspace{19.5bp}
\begingroup
\fontsize{10}{11.05bp}\selectfont
\begin{multicols}{2}
\raggedcolumns
GRID is a student-led transient gamma-ray monitoring constellation \citep{Wen2019,Gao2022}; GRID-11B detected GRB~261010A \citep{Yu2026GCN45940} at the trigger time $T_0 =$ 2026 October 10, 19:17:03 UTC.
Given the limited effective area and sensitivity of GRID-11B, only the main emission episode of GRB~261010A was detected with GRID-11B, while the fainter extended emission reported by other instruments was not. We therefore restrict the analysis presented here to the main episode, including the duration, temporal and spectral properties.

The timing and spectral analysis is performed in the 100--2000~keV band. Linear-background subtraction gives $T_{90}=5.68^{+1.04}_{-0.32}$~s (Fig.~\ref{fig:overview}a). We obtain an amplitude parameter $f=1.76^{+0.14}_{-0.08}$ \citep{Zhang2018} and measure an observer-frame minimum variability timescale of $\Delta t_{\rm min}=51.6^{+10.5}_{-9.8}$~ms \citep{Wang2025,Scargle2013}. The quoted timing uncertainties are at the 68.27\% confidence interval.

Fitting the 100--2000~keV spectrum over $T_{90}$ with a cutoff power-law (CPL) model, we obtain $\alpha=-1.02^{+0.39}_{-0.33}$ and $E_{\rm p}=1089^{+171}_{-80}$~keV (Fig.~\ref{fig:overview}b), with PGSTAT/d.o.f.\ $=168.9/135$. The best-fit model gives a 100--2000~keV flux of $7.83^{+0.45}_{-0.78}\times10^{-7}$~erg~cm$^{-2}$~s$^{-1}$ and a fluence of $4.44^{+0.12}_{-0.23}\times10^{-6}$~erg~cm$^{-2}$.
The rest-frame 1--$10^4$~keV isotropic-equivalent energy is $E_{\rm iso}=6.60^{+0.34}_{-0.45}\times10^{49}$~erg and the rest-frame spectral peak energy is $E_{\rm p,z}=1165^{+183}_{-86}$~keV given a redshift $z=0.070$ \citep{Corcoran2026GCN45912}.
\par\columnbreak
\noindent For an alternative candidate host with a redshift $z=0.785$ \citep{CastroTirado2026GCN45928}, the corresponding values are $E_{\rm iso}=9.50^{+1.13}_{-1.36}\times10^{51}$~erg and $E_{\rm p,z}=1944^{+305}_{-143}$~keV. These estimates include the cosmological $k$-correction \citep{Bloom2001} and assume standard cosmology with $H_0=70$~km~s$^{-1}$~Mpc$^{-1}$, $\Omega_{\rm m}=0.3$, and $\Omega_\Lambda=0.7$.

Despite its long duration, the MVT lies at the short-timescale edge of long GRBs and is typical of short GRBs \citep{Camisasca2023,Wang2025}. Its $f=1.76$ falls in the long/short overlap \citep{Lu2014}. Using $\varepsilon=(E_{\rm iso}/10^{52}\,\mathrm{erg})/(E_{\rm p,z}/100\,\mathrm{keV})^{5/3}$ \citep{Lu2010}, we obtain $\log_{10}\varepsilon=-3.96^{+0.42}_{-0.36}$ and $-2.17^{+0.13}_{-0.18}$ for $z=0.070$ and $0.785$, respectively, both in the low-$\varepsilon$, Type~I-like regime ($\varepsilon<0.03$). Finally, its position relative to the Amati relation is broadly consistent with the Type~I region (Fig.~\ref{fig:overview}c).

The combination of a long duration and Type-I-like prompt-emission properties makes GRB~261010A qualitatively reminiscent of GRB~211211A \citep{Rastinejad2022}, for which an associated kilonova supports a compact-object merger origin. This parallel motivates further investigation of a possible merger origin for GRB~261010A. Its distinctive combination of temporal and spectral properties makes it a valuable case for exploring the diversity of GRB prompt emission and the connection between observational classification and physical progenitors.
\end{multicols}
\endgroup

\vspace{18bp}
\noindent\begin{minipage}{\textwidth}
\centering
\includegraphics[width=\linewidth]{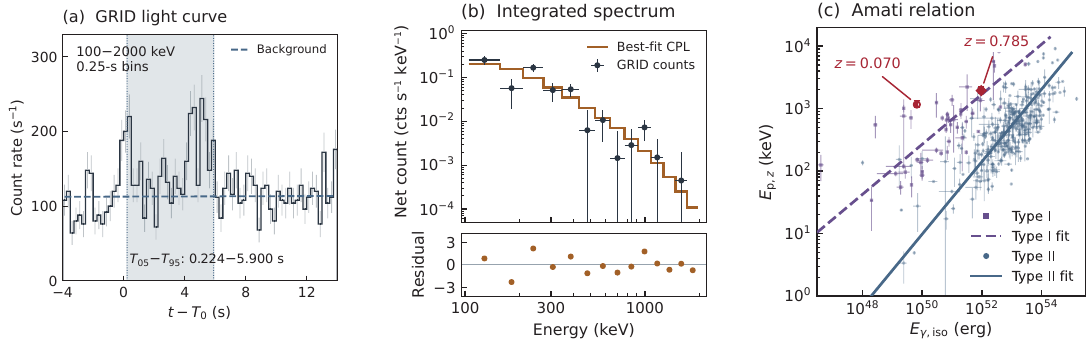}
\captionsetup{font=published}
\captionof{figure}{Observational results of GRB~261010A with GRID-11B. (a) GRID-11B 100--2000~keV light curve; shading indicates $T_{05}$--$T_{95}$ in the same band. (b) Re-binned net counts, response-folded best-fitting CPL, and standardized residuals. (c) GRB~261010A and the Amati relation \citep{Amati2006}. Purple squares and blue circles denote Type~I and Type~II GRBs in the comparison compilation, respectively; the corresponding reference relations are shown as dashed and solid curves, without intrinsic-scatter bands. The red open circle and diamond represent the candidate-host redshifts $z=0.070$ and $z=0.785$, respectively.}\label{fig:overview}
\end{minipage}\par
\vfill
{\raggedleft\sizebp{9}{10}\pageid\par}
\vspace{12bp}
{\centering\sizebp{6}{10}This article is published under a \href{https://creativecommons.org/licenses/by/4.0/}{\textcolor{black}{Creative Commons Attribution 4.0 International (CC BY 4.0) license}}.\par
The authors retain copyright and grant the journal right of first publication. Users are free to copy, distribute, transmit,\par
and adapt the work, provided appropriate credit is given to the original authors and the source is cited.\par}
\newpage

{\centering\sizebp{9}{10.5}\journal\par}
\vspace{17bp}
\begingroup
\setlength{\bibsep}{0pt}
\setlength{\bibhang}{0pt}
\renewcommand{\bibfont}{\sizebp{8}{9}}
\renewcommand{\bibsection}{\noindent{\sffamily\bfseries\sizebp{10.7}{12}References}\par}
\hypersetup{urlcolor=magenta}
\begin{multicols}{2}
\raggedcolumns

\end{multicols}
\endgroup

\vspace{40bp}
{\centering\sizebp{11}{14}
Chenyu Wang$^{1\star}$\orcidlink{0009-0007-9068-2752},
Longhao Li$^{2}$\orcidlink{0009-0009-8633-7280},
Xiaofan Pan$^{2}$\orcidlink{0000-0002-7439-6621},
Hua Feng$^{3}$\orcidlink{0000-0001-7584-6236},
Ming Zeng$^{2}$\orcidlink{0000-0001-9717-1751}
and Zihan Yu$^{2}$\par}
\vspace{14bp}
{\sizebp{9}{10}
\leftskip=23bp\relax
\noindent$^{1}$\quad Department of Astronomy, Tsinghua University, Beijing 100084, China\par
\noindent$^{2}$\quad Department of Engineering Physics, Tsinghua University, Beijing 100084, China\par
\noindent$^{3}$\quad State Key Laboratory of Particle Astrophysics, Institute of High Energy Physics, Chinese Academy of Sciences, Beijing 100049, China\par}
\vfill
{\sizebp{9}{10.5}
\noindent\rule{102bp}{0.4pt}\par
\noindent\hspace{7bp}$\star$\quad Corresponding author: \href{mailto:wang-cy22@mails.tsinghua.edu.cn}{wang-cy22@mails.tsinghua.edu.cn}\par
\noindent Received: 4 November 2026; Revised: 6 November 2026; Accepted: 8 November 2026; Published online: 16 November 2026\par
\vspace{10bp}
\noindent\pageid\par}
\end{document}